\documentclass[3p,onecolumn,12pt,times,longtitle]{elsarticle}

\usepackage{amsmath}
\usepackage{amssymb}
\usepackage{color}
\usepackage{url}
\usepackage{fleqn}

\def\d{{\rm d}}
\def\un#1{\,{\rm #1}}
\def\ung#1{\quad[{\rm #1}]}
\def\unt#1{[{\rm #1}]}
\def\e{{\rm e}}

\def\T{{\rm T}}

\def\mat#1{\mathsf{#1}}
\def\etal{et al.}

\setbox123\hbox{$0$}
\setbox124\hbox{$.$}
\def\S{\hbox to\wd123{\hss}}
\def\.{\hbox to\wd124{\hss}}

\def\Name#1{\textsc{#1}, }
\def\REVIEW#1#2#3#4{{\it #1} {\bf #2} (#3) #4}

\journal{Nuclear Physics B}

\begin{document}

\begin{frontmatter}

\title{Evidence for non-exponential elastic proton-proton differential cross-section at low $|\hbox{t}|$ and $\sqrt{\hbox{s}}$ = 8 TeV by TOTEM}


\def\AddAuthor#1#2#3#4{%
	\def\Name{#1}%
	\def\PriAf{#2}%
	\def\SecAf{#3}%
	\def\ExtAf{#4}%
	\def\empty{}%
	
	\ifx\PriAf\empty
		\author{#1\fnref{#4}}%
	\else
		\ifx\SecAf\empty
			\ifx\ExtAf\empty
				\author[#2]{#1}%
			\else
				\author[#2]{#1\fnref{#4}}%
			\fi
		\else
			\ifx\ExtAf\empty
				\author[#2,#3]{#1}%
			\else
				\author[#2,#3]{#1\fnref{#4}}%
			\fi
		\fi
	\fi
}


\def\AddCorrespondingAuthor#1#2#3#4#5#6{%
	\def\Name{#1}%
	\def\PriAf{#2}%
	\def\SecAf{#3}%
	\def\ExtAf{#4}%
	\def\empty{}%
	
	\ifx\PriAf\empty
		\author{#1\fnref{#4}\corref{CA}}%
	\else
		\ifx\SecAf\empty
			\ifx\ExtAf\empty
				\author[#2]{#1\corref{CA}}%
			\else
				\author[#2]{#1\fnref{#4}\corref{CA}}%
			\fi
		\else
			\ifx\ExtAf\empty
				\author[#2,#3]{#1\corref{CA}}%
			\else
				\author[#2,#3]{#1\fnref{#4}\corref{CA}}%
			\fi
		\fi
	\fi

	\ead{#6}%
	\cortext[CA]{#5}%
}


\def\AddInstitute#1#2{%
	\address[#1]{#2}%
}


\def\AddExternalInstitute#1#2{%
	\fntext[#1]{#2}%
}


\author{TOTEM Collaboration}
\address{\vspace*{-3mm}}
\def\DeclareAuthors{%
	\AddAuthor{G.~Antchev}{}{}{a}%
	\AddAuthor{P.~Aspell}{9}{}{}%
	\AddAuthor{I.~Atanassov}{}{}{a}%
	\AddAuthor{V.~Avati}{8}{9}{}%
	\AddAuthor{J.~Baechler}{9}{}{}%
	\AddAuthor{V.~Berardi}{5b}{5a}{}%
	\AddAuthor{M.~Berretti}{9}{7b}{}%
	\AddAuthor{E.~Bossini}{7b}{}{}%
	\AddAuthor{U.~Bottigli}{7b}{}{}%
	\AddAuthor{M.~Bozzo}{6a}{6b}{}%
	\AddAuthor{P.~Broul\'{i}m}{1a}{}{}%
	\AddAuthor{A.~Buzzo}{6a}{}{}%
	\AddAuthor{F.~S.~Cafagna}{5a}{}{}%
	\AddAuthor{C.~E.~Campanella}{5c}{5a}{}%
	\AddAuthor{M.~G.~Catanesi}{5a}{}{}%
	\AddAuthor{M.~Csan\'{a}d}{4a}{}{b}%
	\AddAuthor{T.~Cs\"{o}rg\H{o}}{4a}{4b}{}%
	\AddAuthor{M.~Deile}{9}{}{}%
	\AddAuthor{F.~De~Leonardis}{5c}{5a}{}%
	\AddAuthor{A.~D'Orazio}{5c}{5a}{}%
	\AddAuthor{M.~Doubek}{1c}{}{}%
	\AddAuthor{K.~Eggert}{10}{}{}%
	\AddAuthor{V.~Eremin}{}{}{e}%
	\AddAuthor{F.~Ferro}{6a}{}{}%
	\AddAuthor{A. Fiergolski}{5a}{}{d}%
	\AddAuthor{F.~Garcia}{3a}{}{}%
	\AddAuthor{V.~Georgiev}{1a}{}{}%
	\AddAuthor{S.~Giani}{9}{}{}%
	\AddAuthor{L.~Grzanka}{8}{}{c}%
	\AddAuthor{C.~Guaragnella}{5c}{5a}{}%
	\AddAuthor{J.~Hammerbauer}{1a}{}{}%
	\AddAuthor{J.~Heino}{3a}{}{}%
	\AddAuthor{A.~Karev}{9}{}{}%
	\AddCorrespondingAuthor{J.~Ka\v{s}par}{1b}{9}{}{Corresponding author at: CERN, Geneva, Switzerland.}{jan.kaspar@cern.ch}%
	\AddAuthor{J.~Kopal}{1b}{}{}%
	\AddAuthor{V.~Kundr\'{a}t}{1b}{}{}%
	\AddAuthor{S.~Lami}{7a}{}{}%
	\AddAuthor{G.~Latino}{7b}{}{}%
	\AddAuthor{R.~Lauhakangas}{3a}{}{}%
	\AddAuthor{R.~Linhart}{1a}{}{}%
	\AddAuthor{E.~Lippmaa}{2}{}{+}%
	\AddAuthor{J.~Lippmaa}{2}{}{}%
	\AddAuthor{M.~V.~Lokaj\'{\i}\v{c}ek}{1b}{}{}%
	\AddAuthor{L.~Losurdo}{7b}{}{}%
	\AddAuthor{M.~Lo~Vetere}{6b}{6a}{+}%
	\AddAuthor{F.~Lucas~Rodr\'{i}guez}{9}{}{}%
	\AddAuthor{M.~Macr\'{\i}}{6a}{}{}%
	\AddAuthor{A.~Mercadante}{5a}{}{}%
	\AddAuthor{N.~Minafra}{9}{5b}{}%
	\AddAuthor{S.~Minutoli}{6a}{}{}%
	\AddAuthor{T.~Naaranoja}{3a}{3b}{}%
	\AddAuthor{F.~Nemes}{4a}{}{b}%
	\AddAuthor{H.~Niewiadomski}{10}{}{}%
	\AddAuthor{E.~Oliveri}{9}{}{}%
	\AddAuthor{F.~Oljemark}{3a}{3b}{}%
	\AddAuthor{R.~Orava}{3a}{3b}{}%
	\AddAuthor{M.~Oriunno}{}{}{f}%
	\AddAuthor{K.~\"{O}sterberg}{3a}{3b}{}%
	\AddAuthor{P.~Palazzi}{9}{}{}%
	\AddAuthor{L.~Palo\v{c}ko}{1a}{}{}%
	\AddAuthor{V.~Passaro}{5c}{5a}{}%
	\AddAuthor{Z.~Peroutka}{1a}{}{}%
	\AddAuthor{V.~Petruzzelli}{5c}{5a}{}%
	\AddAuthor{T.~Politi}{5c}{5a}{}%
	\AddAuthor{J.~Proch\'{a}zka}{1b}{}{}%
	\AddAuthor{F.~Prudenzano}{5c}{5a}{}%
	\AddAuthor{M.~Quinto}{5a}{5b}{}%
	\AddAuthor{E.~Radermacher}{9}{}{}%
	\AddAuthor{E.~Radicioni}{5a}{}{}%
	\AddAuthor{F.~Ravotti}{9}{}{}%
	\AddAuthor{E.~Robutti}{6a}{}{}%
	\AddAuthor{L.~Ropelewski}{9}{}{}%
	\AddAuthor{G.~Ruggiero}{9}{}{}%
	\AddAuthor{H.~Saarikko}{3a}{3b}{}%
	\AddAuthor{A.~Scribano}{7a}{}{}%
	\AddAuthor{J.~Smajek}{9}{}{}%
	\AddAuthor{W.~Snoeys}{9}{}{}%
	\AddAuthor{T.~Sodzawiczny}{9}{}{}%
	\AddAuthor{J.~Sziklai}{4a}{}{}%
	\AddAuthor{C.~Taylor}{10}{}{}%
	\AddAuthor{N.~Turini}{7b}{}{}%
	\AddAuthor{V.~Vacek}{1c}{}{}%
	\AddAuthor{J.~Welti}{3a}{3b}{}%
	\AddAuthor{P.~Wyszkowski}{8}{}{}%
	\AddAuthor{K.~Zielinski}{8}{}{}%
}


\def\DeclareInstitutes{%
	\AddInstitute{1a}{University of West Bohemia, Pilsen, Czech Republic}
	\AddInstitute{1b}{Institute of Physics of the Academy of Sciences of the Czech Republic, Praha, Czech Republic}
	\AddInstitute{1c}{Czech Technical University, Praha, Czech Republic}
	\AddInstitute{2}{National Institute of Chemical Physics and Biophysics NICPB, Tallinn, Estonia}
	\AddInstitute{3a}{Helsinki Institute of Physics, Helsinki, Finland}
	\AddInstitute{3b}{Department of Physics, University of Helsinki, Helsinki, Finland}
	\AddInstitute{4a}{Wigner Research Centre for Physics, RMKI, Budapest, Hungary}
	\AddInstitute{4b}{KRF University College, Gy\"ongy\"os, Hungary}
	\AddInstitute{5a}{INFN Sezione di Bari, Bari, Italy}
	\AddInstitute{5b}{Dipartimento Interateneo di Fisica di Bari, Bari, Italy}
	\AddInstitute{5c}{Dipartimento di Ingegneria Elettrica e dell'Informazione - Politecnico di Bari, Bari, Italy}
	\AddInstitute{6a}{INFN Sezione di Genova, Genova, Italy}
	\AddInstitute{6b}{Universit\`{a} degli Studi di Genova, Italy}
	\AddInstitute{7a}{INFN Sezione di Pisa, Pisa, Italy}
	\AddInstitute{7b}{Universit\`{a} degli Studi di Siena and Gruppo Collegato INFN di Siena, Siena, Italy}
	\AddInstitute{8}{AGH University of Science and Technology, Krakow, Poland}
	\AddInstitute{9}{CERN, Geneva, Switzerland}
	\AddInstitute{10}{Case Western Reserve University, Dept.~of Physics, Cleveland, OH, USA}
}

	
\def\DeclareExternalInstitutes{%
	\AddExternalInstitute{a}{INRNE-BAS, Institute for Nucl. Research and Nucl. Energy, Bulgarian Academy of Sciences, Sofia, Bulgaria.}
	\AddExternalInstitute{b}{Department of Atomic Physics, ELTE University, Budapest, Hungary.}
	\AddExternalInstitute{c}{Institute of Nuclear Physics, Polish Academy of Science, Krakow, Poland.}
	\AddExternalInstitute{d}{Warsaw University of Technology, Warsaw, Poland.}
	\AddExternalInstitute{e}{Ioffe Physical - Technical Institute of Russian Academy of Sciences, St.~Petersburg, Russian Federation.}
	\AddExternalInstitute{f}{SLAC National Accelerator Laboratory, Stanford, CA, USA.}
	\AddExternalInstitute{+}{Deceased.}
}

\DeclareAuthors
\DeclareInstitutes
\DeclareExternalInstitutes

\date{\today}

\begin{abstract}
The TOTEM experiment has made a precise measurement of the elastic 
proton-proton differential cross-section at the centre-of-mass energy 
$\sqrt s = 8\un{TeV}$ based on a high-statistics data sample obtained with 
the $\beta^* = 90\un{m}$ optics. 
Both the statistical and systematic uncertainties remain below $1\un{\%}$, except for the $t$-independent contribution from the overall normalisation. This unprecedented precision allows to exclude a purely exponential differential cross-section in the range of four-momentum transfer squared $0.027 < |t| < 0.2\un{GeV^2}$ with a significance greater than $7\un{\sigma}$. Two extended parametrisations, with quadratic and cubic polynomials in the exponent, are shown to be well compatible with the data. Using them for the differential cross-section extrapolation to $t=0$, and further applying the optical theorem, yields total cross-section estimates of $(101.5 \pm 2.1)\un{mb}$ and $(101.9 \pm 2.1)\un{mb}$, respectively, in agreement with previous TOTEM measurements.

\vbox to 0pt{%
	\vskip-73mm
	\centerline{\footnotesize This article is dedicated to the memory of Prof.~E.~Lippmaa and Prof.~M.~Lo Vetere who passed away recently}%
	\vss
}
\end{abstract}


\end{frontmatter}


\section{Introduction}
The differential cross-section $\d \sigma/\d t$ of hadronic proton-(anti)proton 
scattering at low $|t|$ has traditionally been parametrised with a simple 
exponential function, ${\rm e}^{-B |t|}$, giving a satisfactory description of 
all past experimental data.
Nonetheless, a few experiments have already reported about hints of
slight deviations from this behaviour. At the ISR, for $\sqrt{s}$ 
between 21.5\,GeV and 52.8\,GeV, elastic pp and partly $\rm\bar{p}$p data have shown a 
change of slope~\cite{plb39,plb115} or have been better parametrised with quadratic 
exponential functions, ${\rm e}^{-B |t|-C t^2}$~\cite{npb141,npb248}. 
At the S$\rm\bar{p}$pS, for 
$\sqrt{s} = 546\,$GeV, a change of slope at $|t| \approx 0.14\,\rm GeV^{2}$ 
has been observed, while the inclusion of a quadratic term in the exponent did
not improve the fit significantly~\cite{plb147}. At the Tevatron~\cite{prl61,prl68,prd50,prd86} no 
deviations from pure exponential functions were observed, except at larger $|t|$ where
the influence of the shoulder ($\sim 0.8\,\rm GeV^{2}$ at 
$\sqrt{s} = 0.546$\,TeV and $\sim 0.6\,\rm GeV^{2}$ at 1.8 and 1.96\,TeV) 
becomes visible.
At the LHC, at 7\,TeV as well as at 8\,TeV, all data published so 
far~\cite{epl96,epl101-el,prl111,alfa} have been
compatible with a pure exponential shape.

This report presents a new data sample of elastic scattering at the energy of $\sqrt s = 8\un{TeV}$. Thanks to its high statistics,
an unprecedented precision has been reached in the region $0.027 \lesssim |t| \lesssim 0.2\un{GeV^2}$. Both the statistical and systematic components of the differential cross-section uncertainty are controlled 
at a level below $1\un{\%}$, except for the overall normalisation 
(Section~\ref{sec:normalisation}). Consequently, the functional form of the cross-section can be strongly constrained, thus having more impact on theoretical model building and, in particular, on the extrapolation to $t=0$ used for total cross-section determination. Neglecting the influence of Coulomb scattering in the observed range, the often used purely exponential extrapolation has been found inadequate, and extended parametrisations are provided, still yielding total cross-section values compatible with the previous TOTEM results~\cite{prl111} at the same energy.

This article is organised as follows. Section~\ref{sec:apparatus} outlines the detector apparatus used for this measurement. Section~\ref{sec:data taking} summarises the data-taking conditions; details on the LHC beam optics are given in Section~\ref{sec:beam optics}. Section~\ref{sec:analysis} describes the data analysis and reconstruction of the differential cross-section. In Section~\ref{sec:results} three parametrisations of the differential cross-section are tested, and from those compatible with the data the total cross-section is derived. The results are summarised in Section~\ref{sec:conclusions}.

\section{Experimental apparatus}
\label{sec:apparatus}

The TOTEM experiment is located at the LHC interaction point (IP) 5 together 
with the CMS experiment. In this article only the Roman Pot (RP) system, the 
sub-detector relevant for elastic scattering measurement, is outlined, 
whereas TOTEM's full experimental apparatus is described
elsewhere~\cite{totem-jinst}. 
Roman Pots are movable beam-pipe
insertions that approach the LHC beam very closely in order to detect particles scattered at very small angles. They are organised in two stations placed symmetrically around the IP: one on the left side (in LHC sector 45), one on the right (sector 56). Each station is formed by two units: near ($214\un{m}$ from the IP) and far ($220\un{m}$). Each unit includes three RPs: one approaching the beam from the top, one from the bottom and one horizontally. Each RP hosts a stack of 10 silicon strip sensors (pitch $66\un{\mu m}$) with a strongly reduced insensitive region at the edge facing the beam (few tens of micrometres). The sensors are equipped with trigger-capable electronics. Since elastic scattering events consist of two anti-parallel protons, the detected events can have two topologies, called diagonals: 45 bottom -- 56 top and 45 top -- 56 bottom.

This report will use a reference frame where $x$ denotes the horizontal axis (pointing out of the LHC ring), $y$ the vertical axis (pointing against gravity) and $z$ the beam axis (in the clockwise direction).

\section{Data taking}
\label{sec:data taking}

The measurement presented here is based on data taken in July 2012, during the LHC fill number 2836 providing protons colliding at the centre-of-mass energy $\sqrt s = 8\un{TeV}$. The vertical RPs were inserted at a distance of $9.5$ times the transverse beam size, $\sigma_{\rm beam}$. Initially two, later three colliding bunch-pairs were used, each with a typical population of $8\cdot10^{10}$ protons, yielding an instantaneous luminosity of about $10^{28}\un{cm^{-2}s^{-1}}$ per bunch. The main trigger required a coincidence between the RPs in both arms, combining the near and far units of a station in \textit{OR} to ensure maximal efficiency. During the about $11\un{h}$ long data-taking, a luminosity of $735\un{\mu b^{-1}}$ was accumulated, giving $7.2\cdot 10^6$ tagged elastic events.


\section{Beam optics}
\label{sec:beam optics}

The beam optics relates the proton state at the IP to its state at the RP location. At the IP, the direction of a proton can be described by the scattering angle $\theta^*$ (with respect to the $z$ axis) and azimuthal angle $\phi^*$ (about the $z$ axis). Alternatively, the horizontal ($x$) and vertical ($y$) projections of the scattering angle can be used:
\begin{equation}
\label{eq:scatt angle}
\theta_x^* = \theta^* \cos\phi^*\ ,\qquad \theta_y^* = \theta^* \sin\phi^*\ .
\end{equation}
A proton emerging from the vertex $(x^*$, $y^*)$ at the angle $(\theta_x^*,\theta_y^*)$ and with momentum  $p (1 +  \xi)$, where $p$ is the nominal initial-state proton momentum, arrives at the RPs in a transverse position
\begin{equation}
\label{eq:prot trans}
	x(z_{\rm RP}) = L_x(z_{\rm RP})\, \theta_x^*\ +\ v_x(z_{\rm RP})\, x^*\ +\ D_x(z_{\rm RP})\, \xi\ ,\quad y(z_{\rm RP}) = L_y(z_{\rm RP})\, \theta_y^*\ +\ v_y(z_{\rm RP})\, y^*\ +\ D_y(z_{\rm RP})\, \xi \quad
\end{equation}
relative to the beam centre. This position is determined by the optical functions: effective length $L_{x,y}(z)$, magnification $v_{x,y}(z)$ and dispersion $D_{x,y}(z)$. 
The relative final-state momentum deviation $\xi$ has the following 
contributions:
\begin{itemize}
\item Beam momentum offsets $\xi_{\rm off}$ relative to the nominal momentum 
and time-dependent variations, $\xi_{\rm var}$, with 
$\sigma(\xi_{\rm off}) \sim 10^{-3}$ and $\sigma(\xi_{\rm var}) \sim 10^{-4}$ 
(see discussion in Section~\ref{sec:beam en unc}).
\item The momentum loss, $\xi_{\rm scatt}$, in diffractive scattering processes.
\end{itemize}
For elastic scattering the dispersion terms, $D_{x,y}\, \xi$, can be ignored: 
\begin{itemize}
\item The protons lose no momentum in elastic collisions 
(i.e. $\xi_{\rm scatt} = 0$).
\item Due to the collinearity of the two elastically scattered protons and 
the symmetry of the optics of the two beams, the effects of 
beam energy deviations ($\xi_{\rm off}$ and $\xi_{\rm var}$) on the 
reconstructed scattering angle (Eq.~(\ref{eq:th t}) in 
Section~\ref{sec:kinematics}) are strongly suppressed. Residual effects from 
optics imperfections have been verified to be negligible compared to all other 
uncertainties.
\end{itemize}

For the reported measurement, a special optics with $\beta^* = 90\un{m}$ was used, with essentially the same characteristics as at $\sqrt s = 7\un{TeV}$~\cite{epl96}, see Table~\ref{tab:optics} for details. In the vertical plane, it features parallel-to-point focussing ($v_y \approx 0$) and large effective length $L_y$. In the horizontal plane, the almost vanishing effective length $L_x$ simplifies the separation of elastic and diffractive events: any sizeable horizontal displacement must be due to a momentum loss $\xi$.

\begin{table}
\caption{
Optical functions for elastic proton transport. The values refer to the right arm; for the left arm the moduli are very similar, but $L_{x}$ and $L_{y}$ have
the opposite sign.
}
\label{tab:optics}
\begin{center}
\vskip-3mm
\begin{tabular}{ccccc}\hline\hline
RP unit & $L_x$ & $v_x$ & $L_y$ & $v_y$ \cr\hline
near & $\phantom{-}2.45\un{m}$  & $-2.17$ & $239\un{m}$ & $0.040$ \cr
far  & $-0.37\un{m}$ & $-1.87$ & $264\un{m}$ & $0.021$ \cr
\hline\hline
\end{tabular}
\end{center}
\end{table}

\section{Analysis}
\label{sec:analysis}

The analysis method is similar to the ones used in the previous publications~\cite{epl101-el,prl111}.
However, a different normalisation approach is used (Section~\ref{sec:normalisation}) that makes all $t$-independent scaling factors irrelevant.

The analysis is presented in two main blocks. Section~\ref{sec:event anal} covers all aspects related to the reconstruction of a single event.
Section~\ref{sec:diff cs} describes the steps of transforming a raw $t$-distribution into the differential cross-section. The $t$-distributions for the two diagonals are analysed separately. After comparison (Section~\ref{sec:cross checks}) they are finally merged (Section~\ref{sec:final data merging}).


\subsection{Event analysis}
\label{sec:event anal}

Event kinematics are determined from the coordinates of track hits in the RPs after proper alignment (see Section~\ref{sec:alignment}), using the LHC optics (see Section~\ref{sec:optics}).


\subsubsection{Kinematics reconstruction}
\label{sec:kinematics}

The scattering angles and vertex position are first determined for each proton (i.e.~from each arm) separately by inverting the proton transport, Eq.~(\ref{eq:prot trans}), assuming $\xi = 0$. The following formulae optimise the robustness against optics imperfections:
\begin{equation}
\label{eq:kin 1a}
	\theta_x^{*\rm L,R} = {v_x^{\rm N} x^{\rm F} - v_x^{\rm F} x^{\rm N}\over v_x^{\rm N} L_x^{\rm F} - v_x^{\rm F} L_x^{\rm N}}\ ,\qquad
	\theta_y^{*\rm L,R} = {1\over 2} \left( {y^{\rm N}\over L_y^{\rm N}} + {y^{\rm F}\over L_y^{\rm F}} \right)\ ,\qquad
	x^{*\rm L,R} = {L_x^{\rm F} x^{\rm N} - L_x^{\rm N} x^{\rm F}\over v_x^{\rm N} L_x^{\rm F} - v_x^{\rm F} L_x^{\rm N} }\ ,
\end{equation}
where the N and F superscripts refer to the near and far units, L and R to the 
left and right arm, respectively. This one-arm reconstruction is used for tagging elastic events, where the left and right arm protons are compared.

Once an event is selected, the information from both arms is merged yielding better angular resolution:
\begin{equation}
\label{eq:kin 2a}
	\theta_x^* = {\theta_x^{*\rm L} + \theta_x^{*\rm R} \over 2}\ ,\qquad
	\theta_y^* = {\theta_y^{*\rm L} + \theta_y^{*\rm R} \over 2}\ .
\end{equation}
Eventually, the full scattering angle and four-momentum transfer squared are calculated as

\begin{equation}
\label{eq:th t}
	\theta^* = \sqrt{{\theta_x^*}^2 + {\theta_y^*}^2}\ ,\qquad
	t = - p^2\, \left({\theta_x^*}^2 + {\theta_y^*}^2\right)\ ,
\end{equation}
where $p$ denotes the beam momentum.


\subsubsection{Alignment}
\label{sec:alignment}

The standard three-step procedure \cite{totem-ijmp} has been applied: beam-based alignment prior to the run (as for LHC collimators) followed by two off-line methods. First, track-based alignment for relative positions among RPs, and second, alignment with elastic events for absolute position with respect to the beam. The final uncertainties per unit (common for top and bottom RPs) are: $2\un{\mu m}$ (horizontal shift), $100\un{\mu m}$ (vertical shift) and $0.2\un{mrad}$ (rotation about the beam axis). Propagated through Eqs.~(\ref{eq:kin 1a}) and (\ref{eq:kin 2a}) to the scattering angles reconstructed from both arms, the shifts lead to uncertainties of $0.8\un{\mu rad}$ (horizontal) and $0.2\un{\mu rad}$ (vertical). The relatively large impact of horizontal misalignment is due to the almost vanishing effective length $L_x$ (cf. Eq.~(\ref{eq:kin 1a})). RP rotations induce a bias in the reconstructed horizontal scattering angle:
\begin{equation}
\label{eq:alig rot bias}
	\theta_x^* \rightarrow \theta_x^* + c \theta_y^*\ ,
\end{equation}
where the proportionality constant $c$ has a mean of 0 and a standard deviation of $0.02$.


\subsubsection{Optics}
\label{sec:optics}
In order to reduce the impact of imperfect optics knowledge, the LHC optics calibration \cite{totem-optics} has been applied. This method uses various RP observables to determine fine corrections to the optical functions presented in Eq.~(\ref{eq:prot trans}).

The residual errors induce a bias in the reconstructed scattering angles:
\begin{equation}
\label{eq:opt bias}
	\theta_x^* \rightarrow (1 + d_x)\, \theta_x^*\ ,\qquad
	\theta_y^* \rightarrow (1 + d_y)\, \theta_y^*\ .
\end{equation}
For the two-arm reconstruction, Eq.~(\ref{eq:kin 2a}), the biases $d_x$ and $d_y$ have uncertainties of $0.21\un{\%}$ and $0.25\un{\%}$, respectively, and a correlation factor of $-0.70$. These estimates include the effects of magnet field harmonics. For evaluating the impact on the $t$-distribution, it is convenient to decompose the correlated biases $d_x$ and $d_y$ into eigenvectors of the covariance matrix:
\begin{equation}
\label{eq:opt bias modes}
\begin{pmatrix} d_x\cr d_y \end{pmatrix} =
	\eta_1 \underbrace{\begin{pmatrix} -0.182\un{\%} \cr +0.235\un{\%} \end{pmatrix}}_{\rm mode\ 1}
	\ +\ \eta_2 \underbrace{\begin{pmatrix} -0.096\un{\%} \cr -0.074\un{\%} \end{pmatrix}}_{\rm mode\ 2}
\end{equation}
normalised such that the factors $\eta_{1,2}$ have unit variance.


\subsubsection{Resolution}
\label{sec:resolution}

\begin{figure}
\begin{center}
\includegraphics{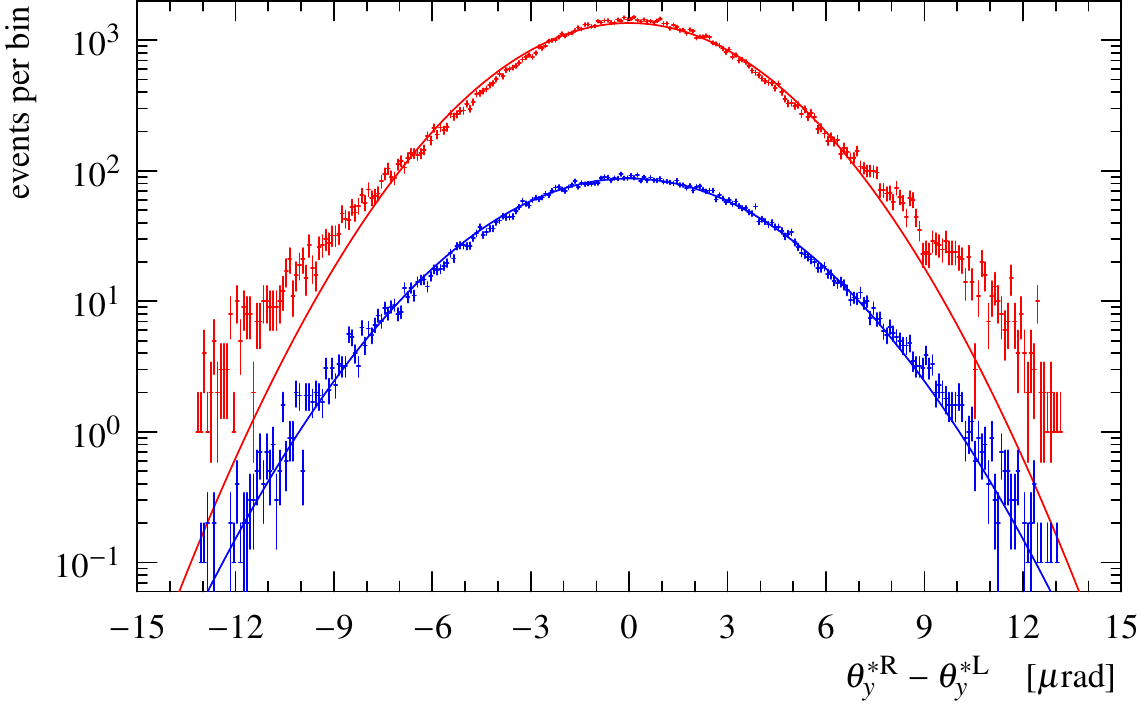}
\vskip-4mm
\caption{%
Difference between vertical scattering angles reconstructed in the right and left arm, for the diagonal 45 bottom - 56 top. Upper graph (red): data from run start ($0.5$ to $1.5\un{h}$ from the beginning of the run). Lower graph (blue): data from run end ($10.5$ to $11.5\un{h}$), scaled by $0.1$. The solid lines represent Gaussian fits.
}
\label{fig:beam divergence}
\end{center}
\end{figure}

Statistical fluctuations in the reconstructed scattering angles are caused by the beam divergence and, in the horizontal projection (due to the small $L_x$), also by the sensor resolution. They are studied by comparing the scattering angles reconstructed from the two arms, in particular through differences $\theta_{x,y}^{*\rm R} - \theta_{x,y}^{*\rm L}$ as illustrated in Figure \ref{fig:beam divergence}. The distributions exhibit small deviations from a Gaussian shape which decrease with time.

\begin{figure}
\begin{center}
\includegraphics{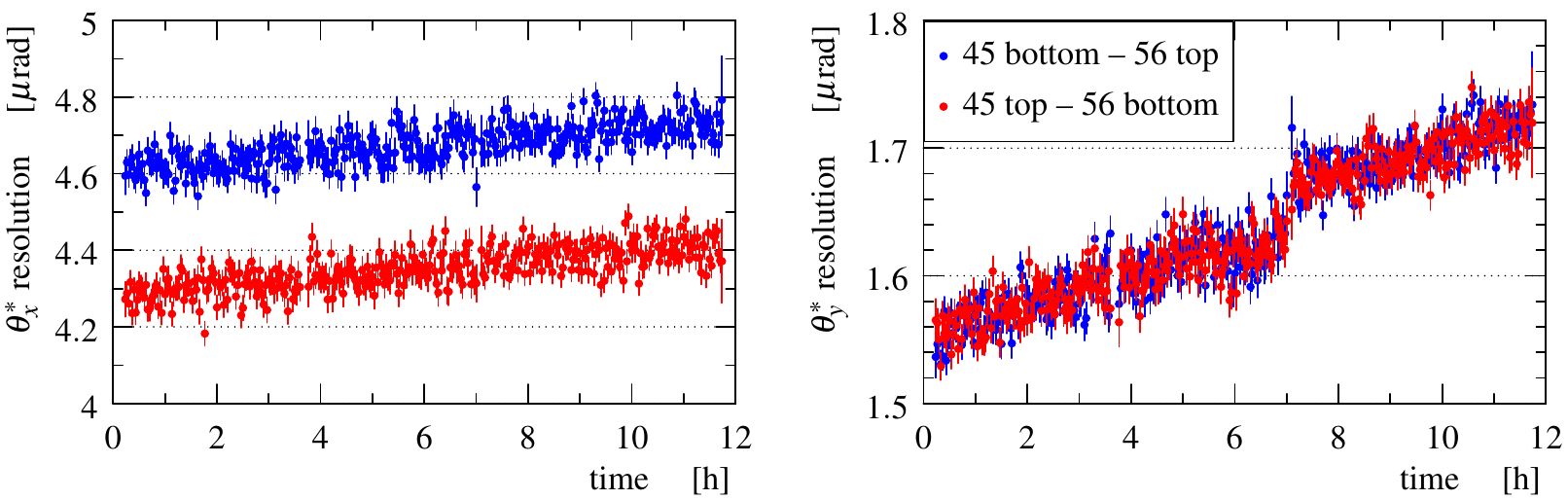}
\vskip-4mm
\caption{%
Angular resolution for the two-arm reconstruction, Eq.~(\ref{eq:kin 2a}), as a function of time (from the beginning of the run). The step in $\theta_y^*$ resolution around $7\un{h}$ is due to inclusion of another colliding bunch-pair with a larger vertical emittance.
}
\label{fig:resolutions}
\end{center}
\end{figure}

Since in good approximation the fluctuations are independent in each arm, the angular resolution for the two-arm reconstruction, Eq.~(\ref{eq:kin 2a}), is given by half of the standard deviation of the $\theta_{x,y}^{*\rm R} - \theta_{x,y}^{*\rm L}$ distributions. As shown in Figure~\ref{fig:resolutions}, the resolution deteriorates slightly with time, which can be expected mainly due to the emittance growth. The small difference in $\theta_x^*$ resolution between the diagonals can be attributed to different RPs, each with slightly different spatial resolution, being involved in the two diagonals.

Measurements of beam emittances~\cite{op-elog} show that the vertical beam divergences of the two beams are equal within a tolerance of about $15\un{\%}$. Exploiting this equality, one can deconvolute the distribution of $\theta_y^{*\rm R} - \theta_y^{*\rm L}$ in order to obtain the beam-divergence distribution, used e.g.~for acceptance corrections discussed in Section~\ref{sec:acc corr} (only required in the vertical plane).

\subsection{Differential cross-section}
\label{sec:diff cs}

For a given $t$ bin, the differential cross-section is evaluated by selecting and counting elastic events:
\begin{equation}
	{\d\sigma\over \d t}(\hbox{bin}) =
		{\cal N}\: {\cal U}({\rm bin})\: {\cal B} \: 
		\frac{1}{\Delta t}
                \sum\limits_{t\, \in\, \text{bin}} {\cal A}(\theta^*, \theta_y^*)\: {\cal E}(\theta_y^*) \: ,
\end{equation}
where $\Delta t$ is the width of the bin, ${\cal N}$ is a normalisation factor, 
and the other symbols stand for various correction factors:
${\cal U}$ for unfolding of resolution effects, ${\cal B}$ for background subtraction, ${\cal A}$ for acceptance correction and ${\cal E}$ for detection and reconstruction efficiency.


\subsubsection{Event tagging}
\label{sec:tagging}

\begin{figure*}
\begin{center}
\includegraphics{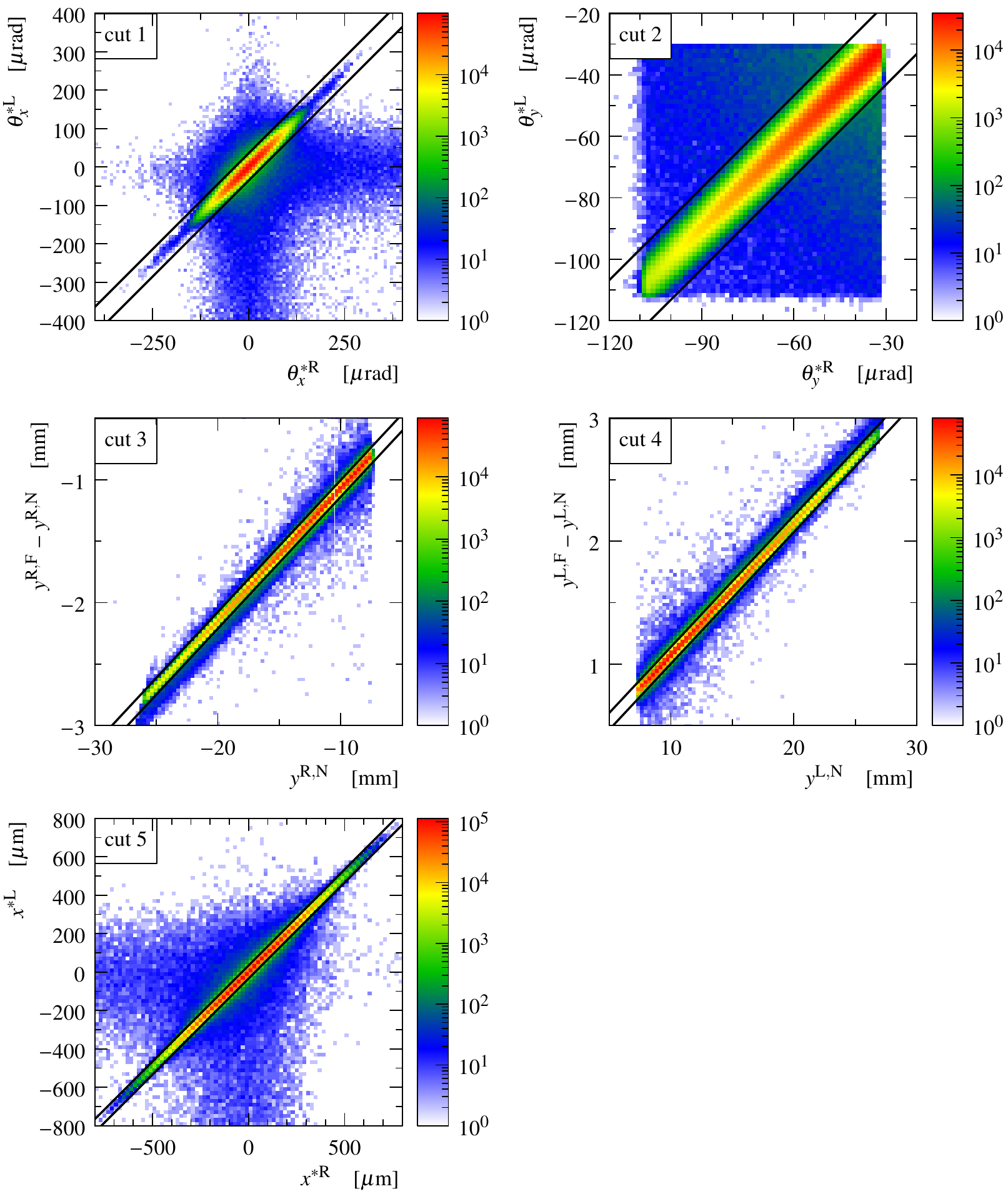}
\caption{%
Correlation plots for the event selection cuts summarised in Table~\ref{tab:cuts}, using all events with diagonal topology 45 top -- 56 bottom. The black solid lines delimit the signal ($\pm 4\un{\sigma}$) region.
}
\label{fig:cuts}
\end{center}
\end{figure*}

The cuts used to select elastic events are summarised in Table~\ref{tab:cuts}. Cuts 1 and 2 require the reconstructed-track collinearity between the left and right arm. Cuts 3 and 4 control the elasticity -- if a proton loses momentum, the vertical position-angle correlation at the RPs is lost. Cut 5 ensures that the two protons come from the same vertex (horizontally). The correlation plots corresponding to these cuts are shown in Figure~\ref{fig:cuts}.

Monte-Carlo simulation suggests that applying all the five cuts at $3\un{\sigma}$ level would lead to a loss of about $2\un{\%}$ of elastic events. Setting the thresholds to $4\un{\sigma}$ yields a tolerable loss of about $0.07\un{\%}$ and therefore the cuts are applied at the $4\un{\sigma}$ level.

The tagging efficiency is studied experimentally by applying the cuts also at the $5\un{\sigma}$ level. This selection yields about $0.5\un{\%}$ more events in every $t$ bin -- thus the inefficiency is irrelevant for this analysis since the overall normalisation is determined from another dataset, see Section~\ref{sec:normalisation}.

\begin{table}
\caption{The elastic selection cuts. The superscripts R and L refer to the right and left arm, N and F correspond to the near and far units, respectively. The constant $\alpha = L_y^{\rm F} / L_y^{\rm N} - 1 \approx 0.11$. The right-most column gives a typical RMS of the cut distribution.
}
\label{tab:cuts}
\begin{center}
\vskip-3mm
\begin{tabular}{ccc}\hline\hline
discriminator & cut quantity & RMS ($\equiv 1\sigma$)\cr\hline
1 & $\theta_x^{*\rm R} - \theta_x^{*\rm L}$				& $9.5\un{\mu rad}$	\cr
2 & $\theta_y^{*\rm R} - \theta_y^{*\rm L}$				& $3.3\un{\mu rad}$	\cr
3 & $\alpha\,y^{\rm R,N} - (y^{\rm R,F} - y^{\rm R,N})$	& $18\un{\mu m}$	\cr
4 & $\alpha\,y^{\rm L,N} - (y^{\rm L,F} - y^{\rm L,N})$	& $18\un{\mu m}$	\cr
5 & $x^{*\rm R} - x^{*\rm L}$							& $8.5\un{\mu m}$ 	\cr\hline\hline
\end{tabular}
\end{center}
\end{table}

\subsubsection{Background}
\label{sec:background}

\begin{figure}
\begin{center}
\includegraphics{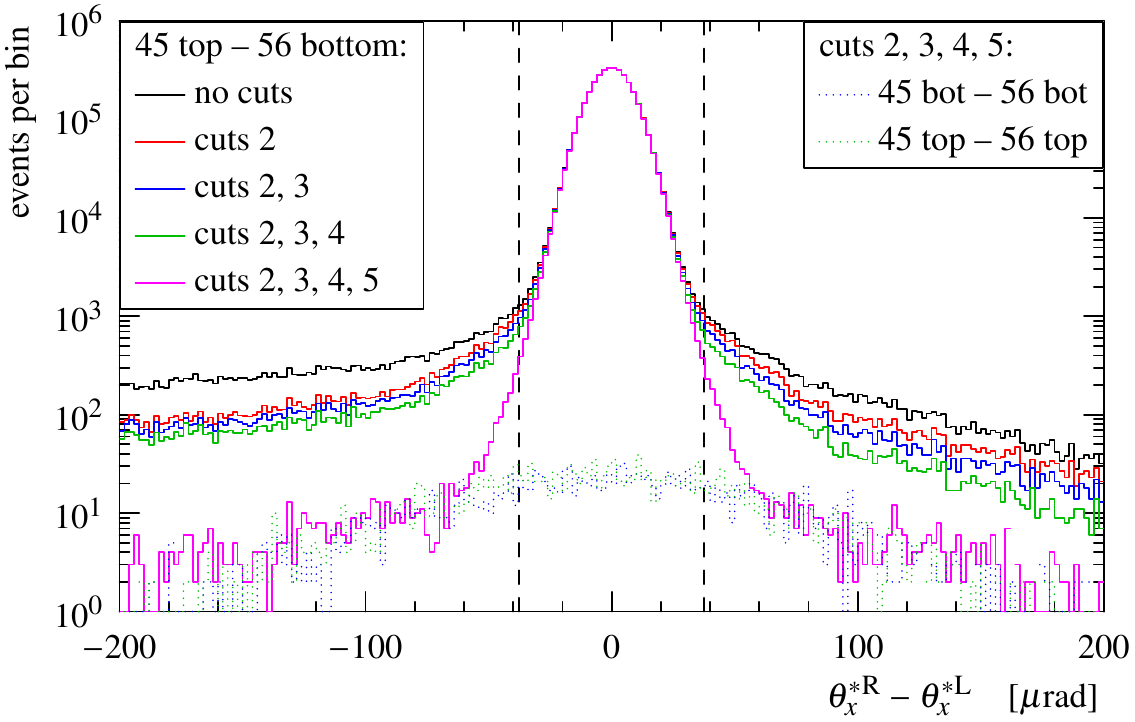}
\caption{%
Distributions of discriminator 1, i.e. the difference between the horizontal scattering angle reconstructed from the right and the left arm. Solid curves: data from diagonal 45 top -- 56 bottom, the different colours correspond to various combinations of the selection cuts (see numbering in Table~\ref{tab:cuts}). Dotted curves: data from non-diagonal RP configurations, obtained by inverting track coordinates in the left arm. The vertical dashed lines represent the boundaries of the signal region ($\pm 4\un{\sigma}$).
}
\label{fig:background}
\end{center}
\end{figure}

Expectable background (i.e.~non-elastic events passing the tagging cuts) may come from central diffraction as well as pile-up of single diffraction and/or beam-halo protons. The background rate is studied by plotting the discriminators from Table~\ref{tab:cuts} under various cut combinations, see an example in Figure~\ref{fig:background}. While the central part (signal) remains essentially constant, the tails (background) are strongly suppressed with increasing number of cuts applied. This interpretation is further supported by the discriminator distributions from non-diagonal RP configurations, see the dotted curves in the figure. While these top -- top or bottom -- bottom configurations cannot contain any elastic signal, they are likely to have a similar share of events causing background to the presented analysis. And indeed, the figure shows a good agreement at the distribution tails. Integrating the non-diagonal curve over the signal region (see the dashed lines in the figure) yields a background estimate of $1 - {\cal B} < 10^{-4}$.

\subsubsection{Acceptance correction}
\label{sec:acc corr}

Two proton detection limitations have been identified: detector coverage (mostly at the edge facing the beam, i.e. relevant for small $|\theta_y^*|$) and LHC apertures ($|\theta_y^*| \approx 100\un{\mu rad}$). The correction accounting for these limitations includes two contributions -- a geometrical correction ${\cal A}_{\rm geom}$ reflecting the fraction of the phase space within the acceptance and a component ${\cal A}_{\rm fluct}$ correcting for fluctuations around the acceptance limits (cuts in $\theta_y^*$):
\begin{equation}
{\cal A}(\theta^*, \theta_y^*) = {\cal A}_{\rm geom}(\theta^*)\ {\cal A}_{\rm fluct}(\theta_y^*)\ .
\end{equation}

\begin{figure}
\begin{center}
\includegraphics{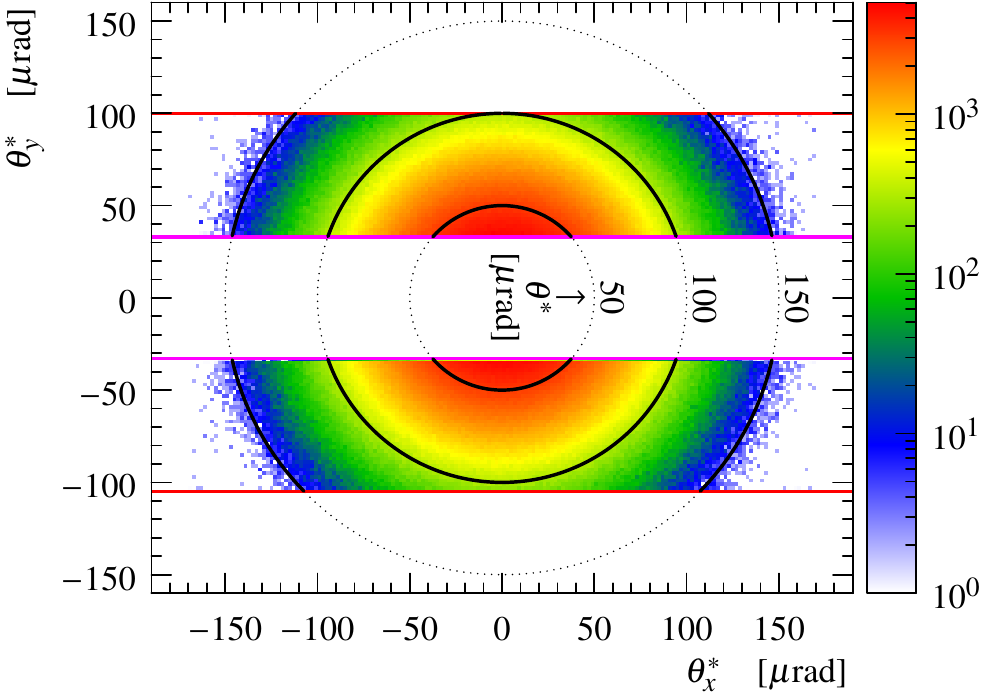}
\vskip-3mm
\caption{%
Distribution of scattering angle projections $\theta_y^*$ vs.~$\theta_x^*$. The upper (lower) part comes from the diagonal 45 bottom -- 56 top (45 top -- 56 bottom). The colour scale gives number of events (after the ${\cal A}_{\rm fluct}$ correction) per bin. The horizontal lines at $\theta_y^* \approx \pm 100\un{\mu rad}$ (red) represent cuts due to the LHC apertures, while those at $\theta_y^* \approx \pm 30 \un{\mu rad}$ (magenta) represent cuts due to the sensor edges. The dotted circles show contours of constant scattering angle $\theta^*$ as indicated in the middle of the plot. The parts of the contours within acceptance are emphasised in thick black. (For interpretation of the references to colour in this figure legend, the reader is referred to the web version of this article.)
}
\label{fig:acceptance principle}
\end{center}
\end{figure}

The calculation of the geometrical correction ${\cal A}_{\rm geom}$ is based on the azimuthal symmetry of elastic scattering, experimentally verified for the data within acceptance. As shown in Figure \ref{fig:acceptance principle}, for a given value of $\theta^*$ the correction is given by:
\begin{equation}
\label{eq:acc geom}
{\cal A_{\rm geom}}(\theta^*) = {
	\hbox{full arc length}\over 
	\hbox{arc length within acceptance}
} \ .
\end{equation}

The correction ${\cal A}_{\rm fluct}$ is calculated analytically from the probability that any of the two elastic protons leaves the region of acceptance due to the beam divergence. The beam divergence distribution is modelled as a Gaussian with the spread determined by the method described in Section~\ref{sec:resolution}. This correction contribution is sizeable only close to the acceptance limits but is kept below $1.5$ by discarding data with larger corrections. The uncertainties are related to the resolution parameters (vertical beam divergence, left-right asymmetry and non-Gaussian shape), and all stay below $0.1\un{\%}$.

\begin{figure}
\begin{center}
\includegraphics{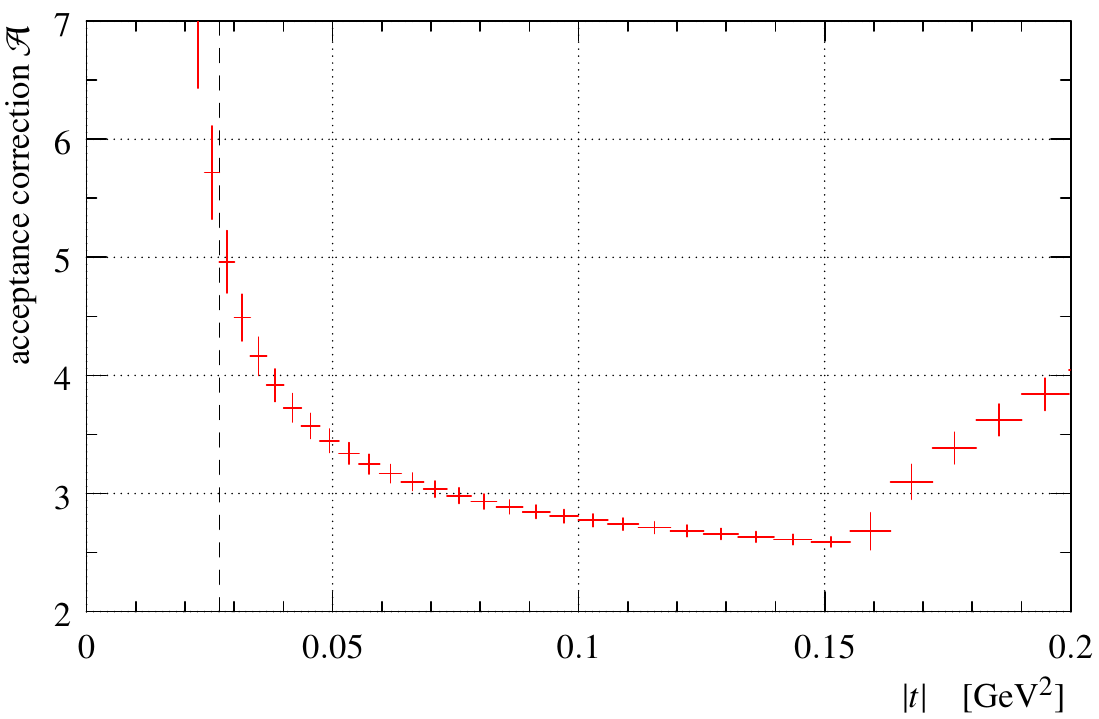}
\vskip-3mm
\caption{%
Full acceptance correction, ${\cal A}$, for diagonal 45 bottom -- 56 top. The points give the mean value per bin, the error bars indicate the standard deviation. The sharp shape change at $|t|\approx 0.16\un{GeV^2}$ is caused by the LHC aperture cuts. The data left of the dashed vertical line are discarded due to excessively large acceptance correction.
}
\label{fig:acceptance result}
\end{center}
\end{figure}

Figure \ref{fig:acceptance result} shows an example for the $t$-dependence of the acceptance correction for one diagonal. Since a single diagonal cannot cover more than half of the phase space, the minimum value of the correction is $2$. As indicated in the figure, data points with too large correction (${\cal A} \gtrsim 5$) are discarded.

\subsubsection{Inefficiency corrections}
\label{sec:ineff corr}

Since the overall normalisation is determined from another dataset (see Section~\ref{sec:normalisation}), any inefficiency correction that does not alter the $t$-distribution shape does not need to be considered in this analysis (trigger, data acquisition and pile-up inefficiency discussed in~\cite{epl101-el,prl111}). The inefficiencies left are related to the inability of a RP to resolve the elastic proton track.

\begin{figure}
\begin{center}
\includegraphics{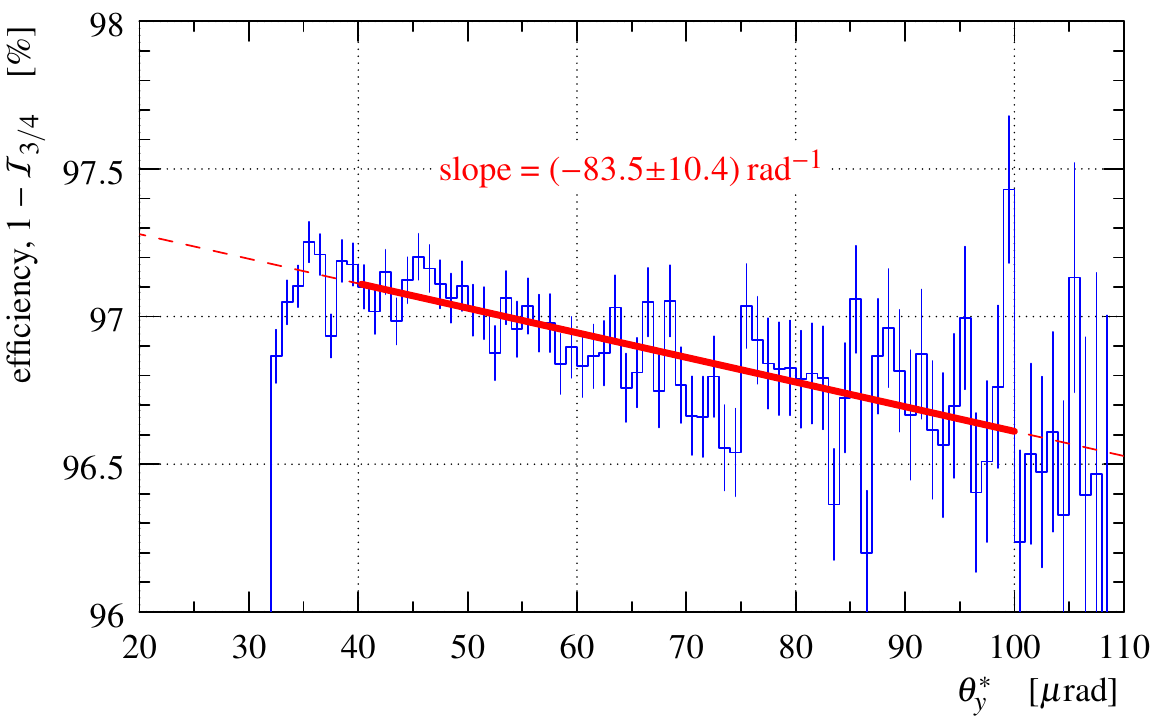}
\vskip-3mm
\caption{%
Single-RP uncorrelated inefficiency for the far top RP in the left arm. The rapid drop at $\theta_y^* \approx 35\un{\mu rad}$ is due to acceptance effects at the sensor edge. The straight (red) lines represent a linear fit of the efficiency dependence on the vertical scattering angle (solid) and its extrapolation to the region affected by acceptance effects (dashed).
}
\label{fig:eff 3/4}
\end{center}
\end{figure}

One such case is when a single RP cannot detect and/or reconstruct a proton track, with no correlation to other RPs. This type of inefficiency, ${\cal I}_{3/4}$, is evaluated by removing the RP from the tagging cuts (Table~\ref{tab:cuts}), repeating the event selection and calculating the fraction of recovered events. A typical example is given in Figure~\ref{fig:eff 3/4}, showing that the efficiency decreases gently with the vertical scattering angle. This dependence stems from the fact that protons with larger $|\theta_y^*|$ hit the RPs further from their edge and therefore the potentially created secondary particles have more chance to induce additional signal in the sensors and thus prevent from resolving the elastic proton track.

Another source of inefficiency are proton interactions in a near RP affecting simultaneously the far RP downstream. The contribution from these near-far correlated inefficiencies, ${\cal I}_{2/4}$, is determined by evaluating the rate of events with high track multiplicity ($\gtrsim$ 5) in both near and far RPs. Events with high track multiplicity simultaneously in a near top and near bottom RP are not counted as such a shower is likely to have started upstream from the RP station and thus unrelated to the elastic proton interacting with detectors. The outcome, ${\cal I}_{2/4} \approx 1.5\un{\%}$, is compatible between left/right arms and top/bottom RP pairs, in addition it compares well to Monte-Carlo simulations (e.g.~section 7.5 in \cite{hubert-thesis}).

The full correction is calculated as
\begin{equation}
\label{efficiency}
	{\cal E}(\theta_y^*) = {1\over 1 - \left( \sum\limits_{i\in \rm RPs} {\cal I}^i_{3/4}(\theta_y^*) + 2 {\cal I}_{2/4} \right) } \ .
\end{equation}
The first term in the parentheses sums the contributions from the four RPs of a diagonal and grows from about $7$ to $10\un{\%}$ from the lowest to the highest $|\theta_y^*|$. The second term amounts to about $3\un{\%}$.



\subsubsection{Unfolding of resolution effects}
\label{sec:unfolding}

\begin{figure}
\begin{center}
\includegraphics{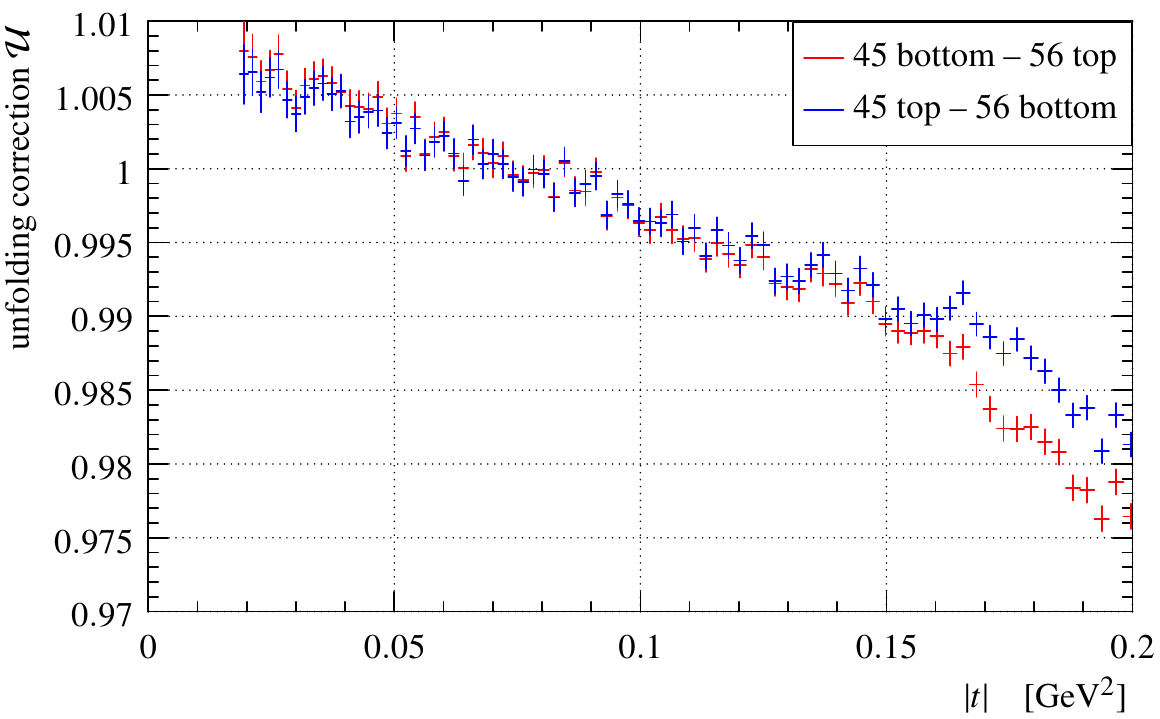}
\vskip-3mm
\caption{%
Unfolding correction as a function of $|t|$. The different shape for $|t| \gtrsim 0.16\un{GeV^2}$ is due to a slightly different position of the LHC aperture cut in the two diagonals.
}
\label{fig:unfolding}
\end{center}
\end{figure}

The correction for resolution effects has been determined by the following iterative procedure. The differential cross-section data are fitted by a smooth curve which serves as an input to a Monte-Carlo simulation using the resolution parameters determined in Section~\ref{sec:resolution}. Making a ratio between simulated histograms with and without smearing effects gives a set of per-bin correction factors. Applying them to the yet uncorrected differential cross-section yields a better estimate of the true $t$-distribution which can be used as input to the next iteration. The iterations stop when the difference between the input and output $t$-distributions becomes negligible, which is typically achieved after the second iteration. Thanks to the good angular resolution (see Section~\ref{sec:resolution}), the final correction is not large, as shown in Figure~\ref{fig:unfolding}.

For the uncertainty estimate, the uncertainties of $\theta_x^*$ and $\theta_y^*$ resolutions (accommodating the full time variation) as well as fit-model dependence have been considered, the first contribution being dominant.

\subsubsection{Normalisation}
\label{sec:normalisation}

The normalisation ${\cal N}$ is determined by requiring the same cross-section integral between $|t| = 0.027$ and $0.083\un{GeV^2}$ as for dataset 1 from \cite{prl111}, where the luminosity-independent calibration was applied. The leading uncertainty of the scaling factor $4.2\un{\%}$ comes from the luminosity-independent method.



\subsubsection{Binning}
\label{sec:binning}

Two binnings have been considered. The ``optimised'' option sets the bin size to $1\un{\sigma}$ of the resolution in $t$. The ``per-mille'' binning is built such that each bin collects about one per-mille of the events.


\subsubsection{Beam energy uncertainty}
\label{sec:beam en unc}

Besides the systematic uncertainties mentioned at the above analysis steps, the uncertainty of the beam momentum needs to be considered when the scattering angles are translated into $t$, see Eq.~(\ref{eq:th t}). The beam momentum at 
$\sqrt{s}=8\,$TeV is derived from the current-to-field calibration functions of the LHC dipole magnets (see Section~4 in~\cite{lhc-note-334}, Section~3.1 in~\cite{cern-ats-2013-040}), yielding a relative momentum uncertainty of $0.07\un{\%}$. Taking into account a further contribution of $0.02\un{\%}$ from quadrupole misalignments, the total relative beam momentum uncertainty amounts to $0.1\un{\%}$, which is the value used in the present analysis.

The precision of this method has been confirmed by direct beam energy measurements at $450\un{GeV}$~\cite{bottura-pac09,fidel}.
Another confirmation is given in \cite{cern-ats-2013-040} (Eq.~(29)), where an alternative beam-momentum measurement based on common proton-ion injections is extrapolated from the injection beam energy of $450\un{GeV}$ to the data-taking energy of $4\un{TeV}$ using the LHC magnetic model. The outcome is consistent with the nominal beam momentum within an uncertainty of $0.1\un{\%}$. When the proton-ion method is directly applied at $4\un{TeV}$, see Eq.~(28) in~\cite{cern-ats-2013-040}, the measurement result is consistent with the above evaluations, but the uncertainty of this method, $0.65\un{\%}$, is larger. 

Finally, energy variations with time during a fill do not exceed $\pm 0.03\un{\%}$ (Section~10 in~\cite{cern-ats-2013-040}) and are hence negligible.


\subsubsection{Propagation of systematic uncertainties}
\label{sec:systematics propagation}

The systematic effects are propagated to the $t$-distribution with help of a Monte-Carlo simulation. A fit of the final differential cross-section data is used to generate the true $t$-distribution. Simultaneously, another $t$-distribution is built, having introduced one of the above mentioned systematic effects at $1\un{\sigma}$ level. The difference between the $t$-distributions gives the systematic effect on the differential cross-section. Formally, this procedure is equivalent to evaluating
\begin{equation}
\label{eq:syst mode}
\delta s_{q}(t) \equiv \frac{\partial(\d\sigma/\d t)}{\partial q} \delta q\ ,
\end{equation}
where $\delta q$ corresponds to $1\un{\sigma}$ bias in the quantity $q$ responsible for a given systematic effect.

The Monte-Carlo simulations show that the combined effect of several systematic errors is well approximated by linear combination of the individual contributions from Eq.~(\ref{eq:syst mode}).


\subsection{Systematic cross-checks}
\label{sec:cross checks}

Compatible results have been obtained from data originating from different bunches, different diagonals and different time periods.

In addition, the complete analysis chain has been applied in two independent analysis implementations, yielding compatible results.


\subsection{Final data merging}
\label{sec:final data merging}

\begin{figure*}
\begin{center}
\includegraphics{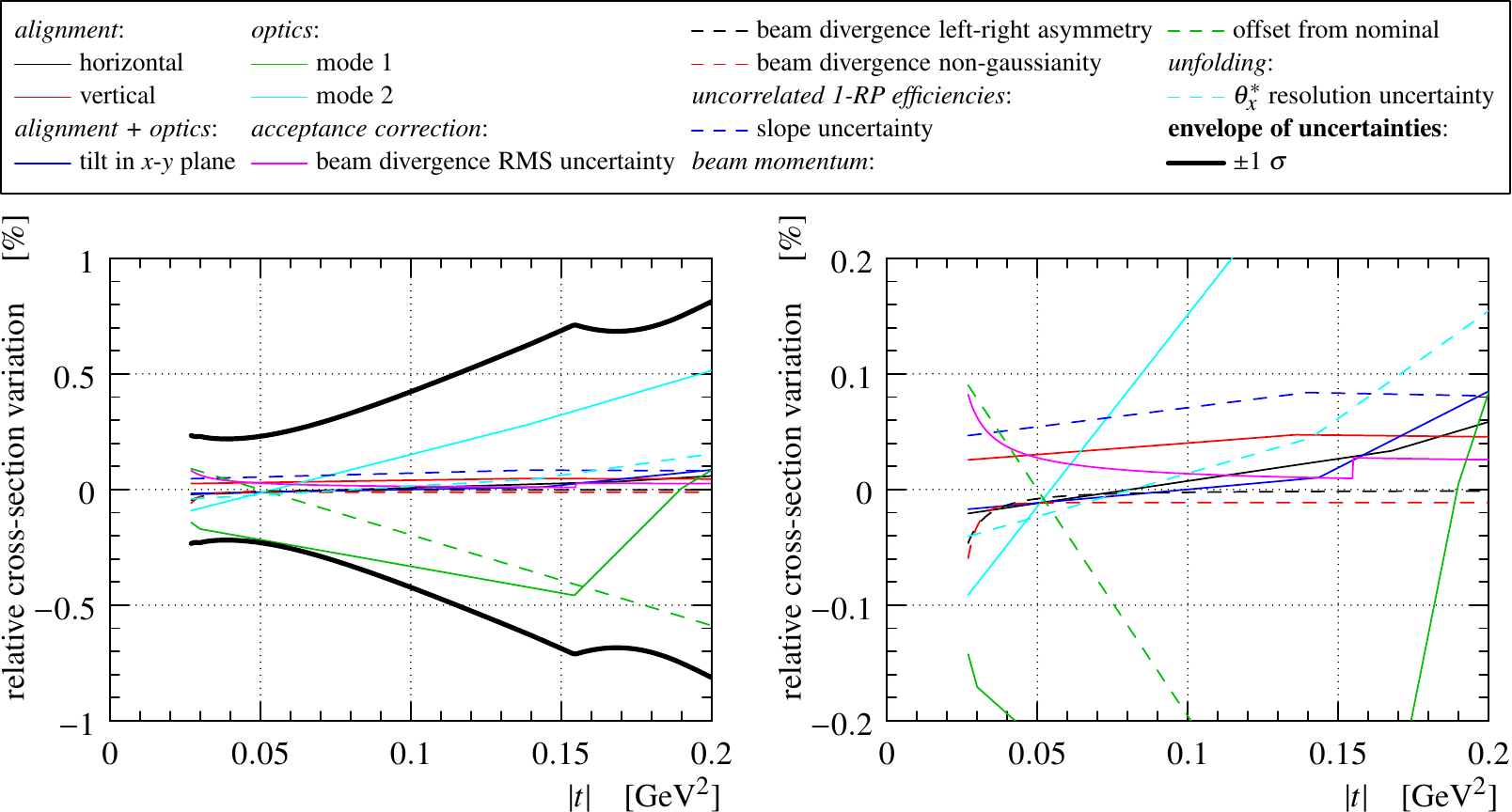}
\vskip-3mm
\caption{%
Impact of $t$-dependent systematic effects on the differential cross-section. Each curve corresponds to a systematic error at $1\un{\sigma}$, cf.~Eq.~(\ref{eq:syst mode}).
The two contributions due to optics correspond to the two vectors in Eq.~(\ref{eq:opt bias modes}).
The envelope is determined by summing all shown contributions in quadrature for each $|t|$ value.
The right plot provides a vertical zoom; note that the envelope is out of scale.
}
\label{fig:syst unc}
\end{center}
\end{figure*}

Finally, the differential cross-section histograms from both diagonals are merged. This is accomplished by a per-bin weighted average, with the weight given by inverse squared statistical uncertainty. The statistical and systematic uncertainties are propagated accordingly. For the systematic ones, the correlation between the diagonals is taken into account. For example, the vertical (mis-)alignment of the RPs within one unit is almost fully correlated, thus the effect on the differential cross-section is opposite in the two diagonals, and consequently its impact is strongly reduced once the diagonals are merged.

The final systematic uncertainties, except the $4.2\un{\%}$ coming from the normalisation, are summarised in Figure~\ref{fig:syst unc} where their impact on the differential cross-section is shown. The leading uncertainties include normalisation, optics imperfections and beam momentum offset. Their effects are quantified in Table~\ref{tab:data}, which can be used to approximate the covariance matrix of systematic uncertainties:
\begin{equation}
\label{eq:covar mat}
\mat V_{ij} = \sum_{q} \delta s_{q}(i)\ \delta s_{q}(j)\: ,
\end{equation}
where $i$ and $j$ are bin indices (row numbers in Table~\ref{tab:data}) and the sum goes over the leading error contributions $q$ (four right-most columns in the table).

\begin{table*}
\caption{%
The elastic differential cross-section as determined in this analysis using the ``optimised'' binning. The three left-most columns describe the bins in $t$. The representative point gives the $t$ value suitable for fitting~\cite{lafferty94}.
The other columns are related to the differential cross-section. The four right-most columns give the leading systematic biases in $\d\sigma/\d t$ for $1\sigma$-shifts in the respective quantities, $\delta s_q$, see Eqs.~(\ref{eq:syst mode}) and (\ref{eq:covar mat}). The two contributions due to optics correspond to the two vectors in Eq.~(\ref{eq:opt bias modes}).
}
\vskip-2mm
\label{tab:data}
\begin{center}
\footnotesize
\setlength{\tabcolsep}{3.2pt}
\begin{tabular}{ccc@{\hskip10pt}ccccccc}
\hline
\hline
\multispan3\hss\vrule width0pt depth4pt height10pt $|t|$ bin $\unt{GeV^2}$\hss & \multispan7\hss $\d\sigma/\d t \ung{mb/GeV^2}$ \hss \cr
\multispan3\hrulefill\hbox to5pt{\hfil} & \multispan7\hrulefill \cr
left & right & represent. & value & statistical     & systematic  & normalisation & optics   & optics   & beam\cr
edge & edge  & point      &       & uncertainty      & uncertainty   &  $\mathcal{N}$     & mode 1   & mode 2   & momentum\cr
\hline
$0.02697$ & $0.03005$ & $0.02850$ & $305.09\S$ & $0.527\S$ & $12.85\S$ & $+12.83\S$ & $-0.479\S$ & $-0.263\S$ & $+0.257\S$ \cr
$0.03005$ & $0.03325$ & $0.03164$ & $287.95\S$ & $0.478\S$ & $12.08\S$ & $+12.06\S$ & $-0.502\S$ & $-0.217\S$ & $+0.206\S$ \cr
$0.03325$ & $0.03658$ & $0.03491$ & $269.24\S$ & $0.436\S$ & $11.32\S$ & $+11.31\S$ & $-0.491\S$ & $-0.174\S$ & $+0.159\S$ \cr
$0.03658$ & $0.04005$ & $0.03831$ & $251.31\S$ & $0.401\S$ & $10.59\S$ & $+10.57\S$ & $-0.478\S$ & $-0.135\S$ & $+0.115\S$ \cr
$0.04005$ & $0.04365$ & $0.04184$ & $235.15\S$ & $0.371\S$ & $\S9.874$ & $+\S9.861$ & $-0.465\S$ & $-0.0981$ & $+0.0750$ \cr
$0.04365$ & $0.04740$ & $0.04551$ & $218.32\S$ & $0.343\S$ & $\S9.185$ & $+\S9.172$ & $-0.451\S$ & $-0.0647$ & $+0.0383$ \cr
$0.04740$ & $0.05129$ & $0.04933$ & $202.64\S$ & $0.318\S$ & $\S8.521$ & $+\S8.509$ & $-0.437\S$ & $-0.0343$ & $+0.0052$ \cr
$0.05129$ & $0.05534$ & $0.05330$ & $187.10\S$ & $0.295\S$ & $\S7.882$ & $+\S7.870$ & $-0.421\S$ & $-0.0070$ & $-0.0244$ \cr
$0.05534$ & $0.05956$ & $0.05743$ & $173.06\S$ & $0.274\S$ & $\S7.270$ & $+\S7.257$ & $-0.405\S$ & $+0.0172$ & $-0.0504$ \cr
$0.05956$ & $0.06394$ & $0.06173$ & $158.77\S$ & $0.255\S$ & $\S6.685$ & $+\S6.672$ & $-0.388\S$ & $+0.0385$ & $-0.0731$ \cr
$0.06394$ & $0.06850$ & $0.06620$ & $144.93\S$ & $0.236\S$ & $\S6.127$ & $+\S6.114$ & $-0.370\S$ & $+0.0569$ & $-0.0925$ \cr
$0.06850$ & $0.07324$ & $0.07085$ & $133.12\S$ & $0.219\S$ & $\S5.597$ & $+\S5.584$ & $-0.352\S$ & $+0.0724$ & $-0.109\S$ \cr
$0.07324$ & $0.07817$ & $0.07568$ & $121.24\S$ & $0.203\S$ & $\S5.096$ & $+\S5.082$ & $-0.334\S$ & $+0.0853$ & $-0.122\S$ \cr
$0.07817$ & $0.08329$ & $0.08071$ & $109.77\S$ & $0.188\S$ & $\S4.623$ & $+\S4.609$ & $-0.316\S$ & $+0.0957$ & $-0.132\S$ \cr
$0.08329$ & $0.08862$ & $0.08593$ & $\S99.077$ & $0.174\S$ & $\S4.179$ & $+\S4.164$ & $-0.297\S$ & $+0.104\S$ & $-0.140\S$ \cr
$0.08862$ & $0.09417$ & $0.09137$ & $\S89.126$ & $0.161\S$ & $\S3.762$ & $+\S3.747$ & $-0.279\S$ & $+0.109\S$ & $-0.145\S$ \cr
$0.09417$ & $0.09994$ & $0.09702$ & $\S79.951$ & $0.148\S$ & $\S3.374$ & $+\S3.359$ & $-0.260\S$ & $+0.113\S$ & $-0.147\S$ \cr
$0.09994$ & $0.10593$ & $0.10290$ & $\S71.614$ & $0.137\S$ & $\S3.014$ & $+\S2.998$ & $-0.242\S$ & $+0.115\S$ & $-0.148\S$ \cr
$0.10593$ & $0.11217$ & $0.10902$ & $\S63.340$ & $0.125\S$ & $\S2.680$ & $+\S2.664$ & $-0.224\S$ & $+0.115\S$ & $-0.147\S$ \cr
$0.11217$ & $0.11866$ & $0.11538$ & $\S56.218$ & $0.115\S$ & $\S2.373$ & $+\S2.357$ & $-0.206\S$ & $+0.114\S$ & $-0.144\S$ \cr
$0.11866$ & $0.12540$ & $0.12199$ & $\S49.404$ & $0.105\S$ & $\S2.092$ & $+\S2.075$ & $-0.189\S$ & $+0.111\S$ & $-0.139\S$ \cr
$0.12540$ & $0.13242$ & $0.12887$ & $\S43.300$ & $0.0961$ & $\S1.835$ & $+\S1.818$ & $-0.173\S$ & $+0.107\S$ & $-0.134\S$ \cr
$0.13242$ & $0.13972$ & $0.13602$ & $\S37.790$ & $0.0876$ & $\S1.601$ & $+\S1.585$ & $-0.157\S$ & $+0.102\S$ & $-0.127\S$ \cr
$0.13972$ & $0.14730$ & $0.14346$ & $\S32.650$ & $0.0795$ & $\S1.391$ & $+\S1.374$ & $-0.142\S$ & $+0.0974$ & $-0.120\S$ \cr
$0.14730$ & $0.15520$ & $0.15120$ & $\S28.113$ & $0.0720$ & $\S1.201$ & $+\S1.185$ & $-0.127\S$ & $+0.0924$ & $-0.112\S$ \cr
$0.15520$ & $0.16340$ & $0.15925$ & $\S24.155$ & $0.0659$ & $\S1.030$ & $+\S1.016$ & $-0.0955$ & $+0.0866$ & $-0.104\S$ \cr
$0.16340$ & $0.17194$ & $0.16761$ & $\S20.645$ & $0.0616$ & $\S0.877$ & $+\S0.866$ & $-0.0590$ & $+0.0804$ & $-0.0951$ \cr
$0.17194$ & $0.18082$ & $0.17632$ & $\S17.486$ & $0.0574$ & $\S0.743$ & $+\S0.733$ & $-0.0302$ & $+0.0739$ & $-0.0865$ \cr
$0.18082$ & $0.19005$ & $0.18537$ & $\S14.679$ & $0.0543$ & $\S0.626$ & $+\S0.617$ & $-0.0081$ & $+0.0673$ & $-0.0780$ \cr
$0.19005$ & $0.19965$ & $0.19478$ & $\S12.291$ & $0.0504$ & $\S0.524$ & $+\S0.515$ & $+0.0052$ & $+0.0606$ & $-0.0697$ \cr
\hline
\hline
\end{tabular}
\end{center}
\end{table*}


\subsection{Statistical uncertainty adjustment}
\label{sec:stat unc adj}

The statistical fluctuations in the differential cross-section using the ``optimised'' binning have been slightly overestimated, whereas the ``per-mille'' binning does not suffer from this problem. One way to demonstrate this is to split the data into groups of consecutive points small enough for a linear function to approximate well the differential cross-section within each group. Then, performing straight-line fits through each group yields on average $\chi^2$ values slightly too low. Alternatively, the issue can be demonstrated as follows. The data sample is divided into several sub-samples corresponding to the same luminosity, and the analysis method described in the earlier sections is repeated for each of these sub-samples. Then, fluctuations of each bin content are determined from the several sub-samples, giving values slightly lower than the uncertainty estimates.

As a remedy, the statistical uncertainties in the ``optimised'' binning have been divided by a factor of $1.176$. This value has been determined by requiring both binnings to give the same value of $\chi^2/\hbox{ndf}$ for fits of $\d\sigma / \d t$ to the fit function in Eq.~(\ref{eq:fit param}) with $N_b = 3$ which has enough flexibility to describe the data.


\section{Results}
\label{sec:results}


\begin{figure}
\begin{center}
\includegraphics{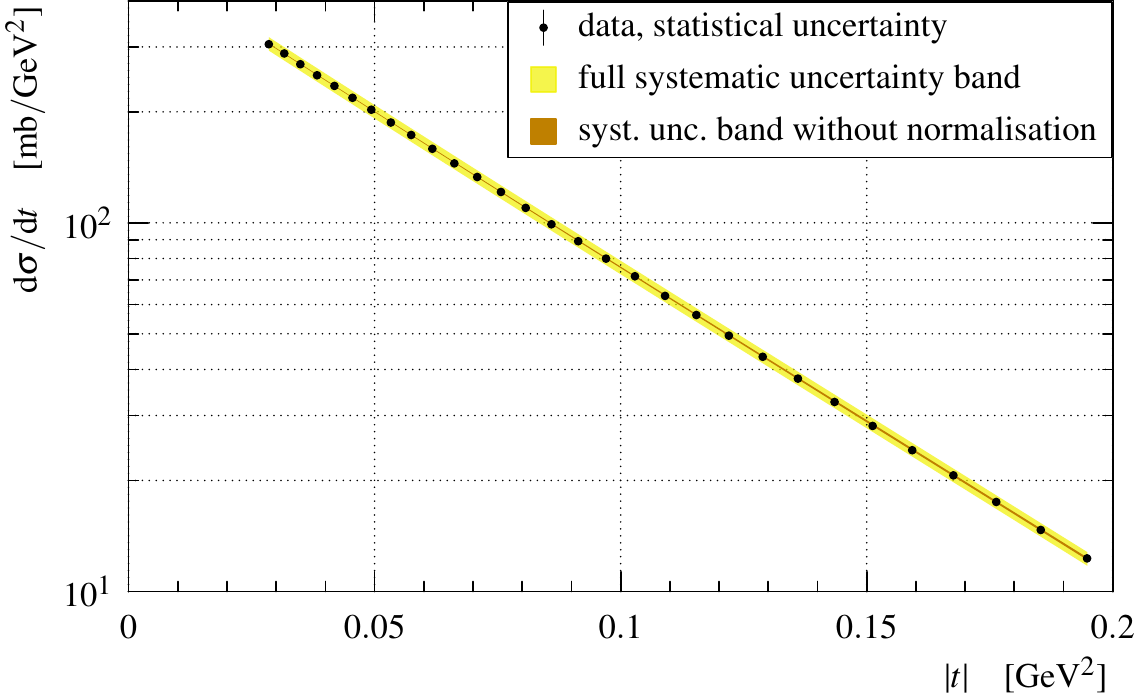}
\caption{%
Differential cross-section using ``optimised'' binning, as given in Table~\ref{tab:data}.
}
\label{fig:data ob}
\end{center}
\end{figure}

\begin{figure*}
\vskip-5mm
\begin{center}
\includegraphics{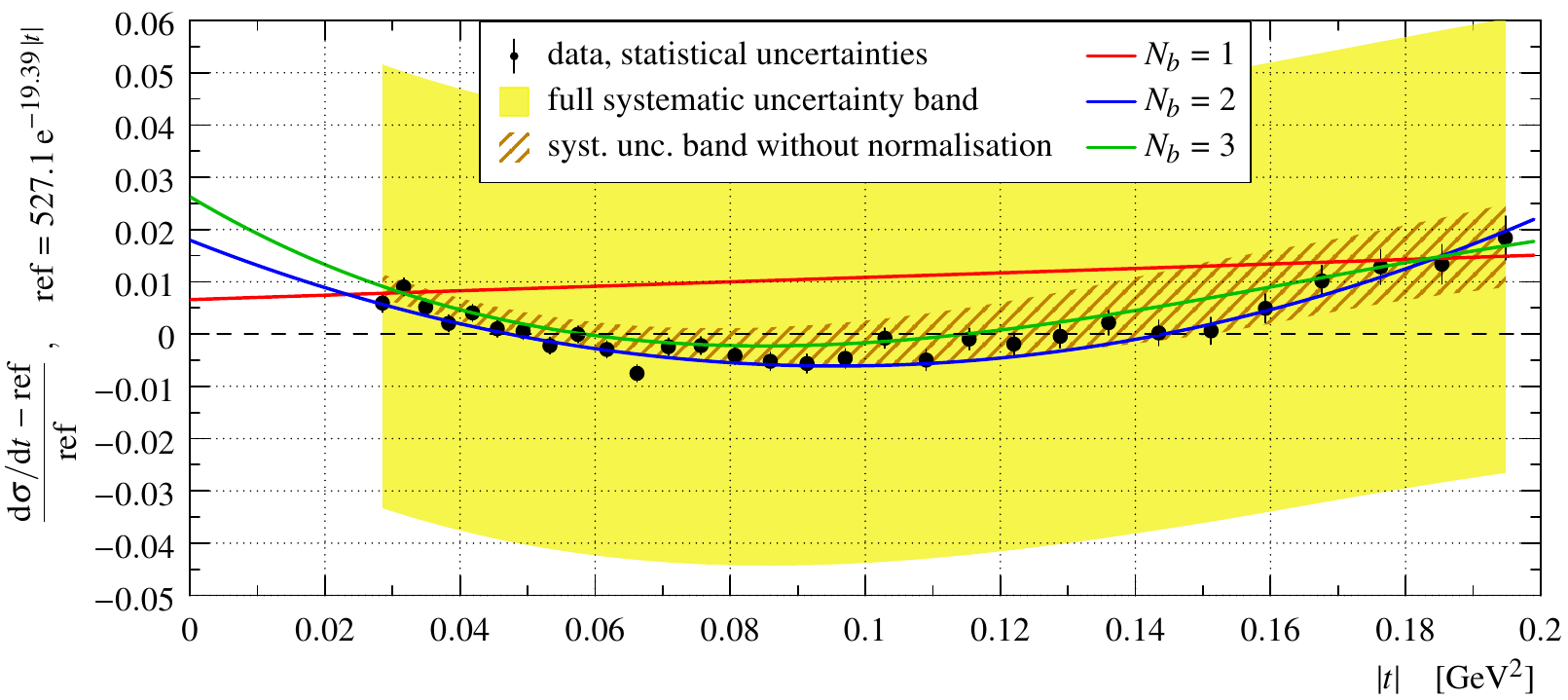}
\vskip-4mm
\caption{%
Differential cross-section using the ``optimised'' binning and plotted as relative difference from a reference exponential (see vertical axis). The black dots represent data points with statistical uncertainty bars. The coloured continuous curves correspond to fits with parametrisation Eq.~(\ref{eq:fit param}) and different numbers of parameters in the exponent. The straight (red) line lies seemingly too high with respect to the data points, which is a consequence of the systematic degrees of freedom included in the fit: some of the effects in Figure~\ref{fig:syst unc} may flatten the distribution which at the same time changes the overall normalisation. The widest error band (yellow) corresponds to the full systematic uncertainty, the hatched (brown) one includes all systematic contributions except the normalisation. Both bands are centred around the fit curve with $N_b = 3$. (For interpretation of the references to colour in this figure legend, the reader is referred to the web version of this article.)
}
\label{fig:data rel ob}
\end{center}
\vskip-2mm
\end{figure*}

The final differential cross-section in the ``optimised'' binning is presented in Table~\ref{tab:data} and \linebreak Figure~\ref{fig:data ob}. In order to visualise small deviations from the leading pure-exponential behaviour, Figure~\ref{fig:data rel ob} shows the relative difference of the cross-section from a reference exponential (pure exponential fit using statistical uncertainties only). This plot immediately suggests a non-exponentiality of the data: pure exponentials would look like (almost) linear functions in this kind of representation.

To study the detailed behaviour of the differential cross-section, a series of fits has been made using the parametrisation:
\begin{equation}
\label{eq:fit param}
{\d\sigma\over\d t}(t) = \left. \d\sigma\over\d t\right|_{t=0} \ \exp\left( \sum\limits_{i = 1}^{N_{b}} b_i\, t^i \right) \ ,
\end{equation}
which includes the pure exponential ($N_b = 1$) and its straight-forward extensions ($N_b = 2, 3$).

The fits have been performed by the standard least-squares method, in particular minimising:
\begin{equation}
\label{eq:chi sq}
	\chi^2 = \Delta^\T \mat V^{-1} \Delta\ ,\qquad
	\Delta_i = \left.{\d\sigma\over \d t}\right|_{{\rm bin}\ i} - {1\over \Delta t_i} \int_{{\rm bin}\ i} f(t)\,\d t\ ,\qquad
	\mat V = \mat V_{\rm stat} + \mat V_{\rm syst}\ ,
\end{equation}
where $\Delta$ is a vector of differences between the differential cross-section data and a fit function $f$, with $\Delta t_i$ representing the width of the $i$-th bin. The covariance matrix $\mat V$ is given by the sum of the statistical component $\mat V_{\rm stat}$ (statistical uncertainty squared from Table~\ref{tab:data} on the diagonal) and the systematic component $\mat V_{\rm syst}$ (see Eq.~(\ref{eq:covar mat})).

The quality of fits is judged on the basis of several measures. The first is the value of $\chi^2$ after minimisation divided by the number of degrees of freedom (ndf). Secondly, the p-value stands for the probability that a $\chi^2$ value greater than the observed one would be drawn from the $\chi^2$ distribution with the given number of degrees of freedom. Finally, significance means the half-width of a central region that needs to be excluded from a normal distribution to get the same integrated probability as the p-value. The significance is expressed in multiples of sigma, the standard deviation of the normal distribution.

Figure~\ref{fig:data rel ob} shows several fits of the differential cross-section with the parametrisation in Eq.~(\ref{eq:fit param}) and different numbers of parameters in the exponent, $N_b$. The corresponding fit quality is given in Table~\ref{tab:fits ob}, indicating that the purely exponential fit ($N_b = 1$) is excluded at $7.2\un{\sigma}$ significance.
The other two fits ($N_b = 2, 3$) are of reasonable quality and can, therefore, be used for a total cross-section estimation with the optical theorem in the form
\begin{equation}
\label{eq:si tot}
\sigma_{\rm tot}^2 = {16\pi\, (\hbar c)^2\over 1 + \rho^2}\, \left. \d\sigma_{\rm el}\over\d t\right|_{t=0}\ ,
\end{equation}
which neglects the effects due to the Coulomb interaction.
Using the COMPETE~\cite{compete} preferred-model extrapolation of $\rho = 0.140\pm 0.007$ yields
\begin{equation}
\label{eq:si tot results}
	\begin{aligned}
		N_b &= 2:\quad \sigma_{\rm tot} = (101.5 \pm 2.1)\un{mb}\ ,\\
		N_b &= 3:\quad \sigma_{\rm tot} = (101.9 \pm 2.1)\un{mb}\ ,\\
	\end{aligned}
\end{equation}
which are well compatible with the previous measurement using the luminosity-independent \linebreak method~\cite{prl111}.

\begin{table}
\caption{%
Details of the fits in Figure~\ref{fig:data rel ob} using parametrisation Eq.~(\ref{eq:fit param}). The matrices give the correlation factors between the fit parameters.
}
\vskip-3mm
\label{tab:fits ob}
\begin{center}
\footnotesize
\begin{tabular}{cccccccc}
\hline
\hline
$N_b$ & $\left . \d\sigma / \d t \right|_{t=0}$ & $b_1$ & $b_2$ & $b_3$ & $\chi^2/\hbox{ndf}$ & p-value & significance\cr
& $\unt{mb/GeV^2}$ & $\unt{GeV^{-2}}$ & $\unt{GeV^{-4}}$ & $\unt{GeV^{-6}}$ & & & \cr
\hline
$1$ & $531 \pm 22$ & $-19.35 \pm 0.06$ & - & -& $117.5/28=4.20$ & $6.2\cdot 10^{-13}$ & $7.20\un{\sigma}$ \cr
        & $ \left(\begin{matrix} +1.00\cr-0.11\cr\end{matrix} \right. $
        & $ \left.\begin{matrix} -0.11\cr+1.00\cr\end{matrix} \right) $
\cr\hline
$2$ & $537 \pm 22$ & $-19.89 \pm 0.08$ & $2.61 \pm 0.30$ & -& $ 29.3/27=1.09$ & $0.35$ & $0.94\un{\sigma}$ \cr
        & $ \left(\begin{matrix} +1.00\cr+0.19\cr-0.34\cr\end{matrix} \right. $
        & $ \begin{matrix} +0.19\cr+1.00\cr-0.76\cr\end{matrix}  $
        & $ \left.\begin{matrix} -0.34\cr-0.76\cr+1.00\cr\end{matrix} \right) $
\cr\hline
$3$ & $541 \pm 22$ & $-20.14 \pm 0.15$ & $5.95 \pm 1.75$ & $-12.0 \pm 6.2$& $ 25.5/26=0.98$ & $0.49$ & $0.69\un{\sigma}$ \cr
        & $ \left(\begin{matrix} +1.00\cr+0.08\cr-0.04\cr-0.02\cr\end{matrix} \right. $
        & $ \begin{matrix} +0.08\cr+1.00\cr-0.90\cr+0.85\cr\end{matrix}  $
        & $ \begin{matrix} -0.04\cr-0.90\cr+1.00\cr-0.99\cr\end{matrix}  $
        & $ \left.\begin{matrix} -0.02\cr+0.85\cr-0.99\cr+1.00\cr\end{matrix} \right) $
\cr
\hline
\hline
\end{tabular}
\end{center}
\end{table}

\begin{figure*}
\begin{center}
\includegraphics{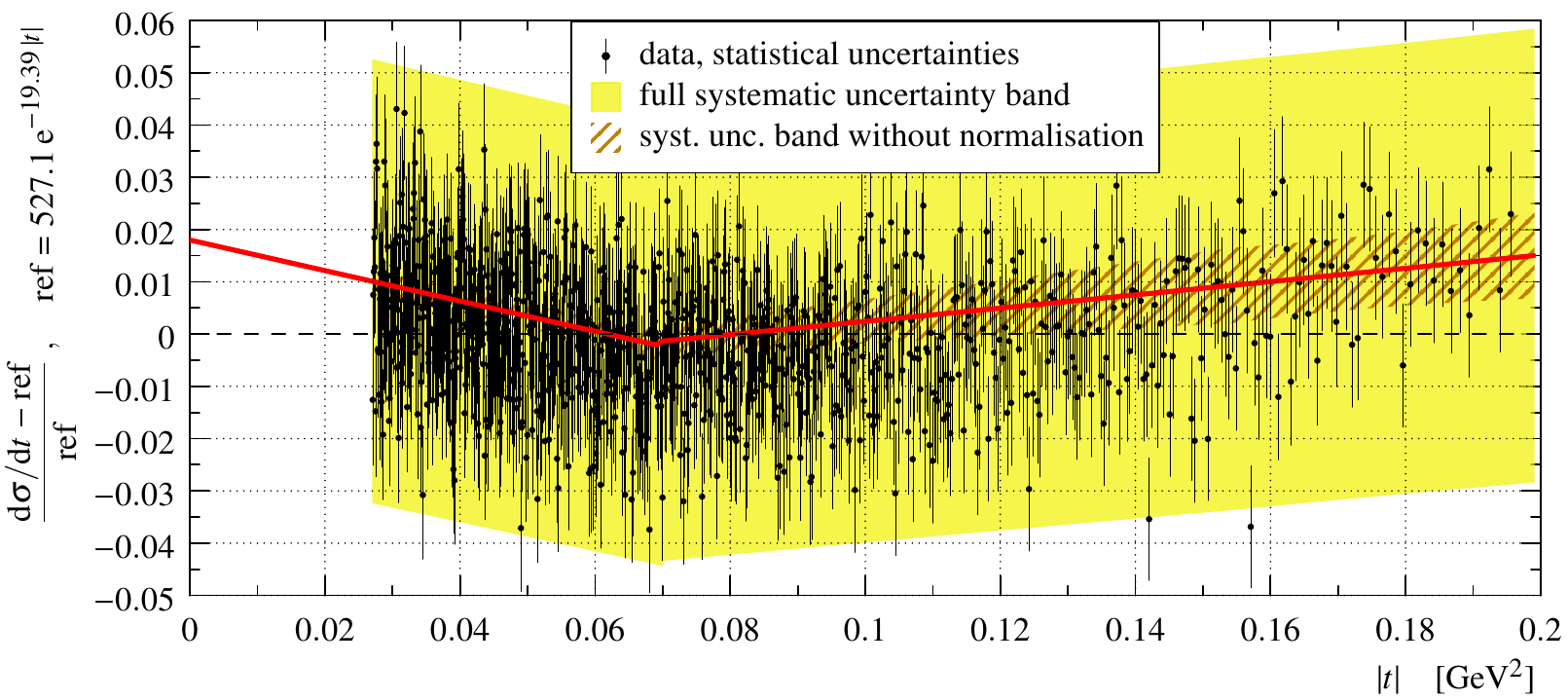}
\vskip-4mm
\caption{%
Differential cross-section using the ``per-mille'' binning and plotted as relative difference from the reference exponential (see vertical axis). The black dots represent data points with statistical uncertainty bars. The red line (with two straight segments) shows pure exponential fits in regions below and above $|t| = 0.07\un{GeV^2}$, see Eq.~(\ref{eq:split fit param}). The widest error band (yellow) corresponds to the full systematic uncertainty, the hatched one (brown) includes all systematic contributions except the normalisation. Both bands are centred around the fit curve. (For interpretation of the references to colour in this figure legend, the reader is referred to the web version of this article.)
}
\label{fig:data rel cpb0.001}
\end{center}
\end{figure*}

The incompatibility between a pure-exponential behaviour and the data with the ``per-mille'' binning can be shown equally well. However, since the number of points is drastically increased, the straight-forward $\chi^2$ test does not have sufficient sensitivity, and a different test is used. Assuming that the data can be described by a pure exponential, the fit parameters should have compatible values for fits over different ranges. Figure~\ref{fig:data rel cpb0.001} shows a fit (minimisation of $\chi^2$ from Eq.~(\ref{eq:chi sq})) with the parametrisation
\begin{equation}
\label{eq:split fit param}
{\d\sigma\over\d t}(t) =
\begin{cases}
a_1\, \e^{b_1 |t|} & |t| < 0.07\un{GeV^2}\\
a_2\, \e^{b_2 |t|} & |t| > 0.07\un{GeV^2}\\
\end{cases}
\end{equation}
giving a reasonable fit quality (p-value of 0.57). The compatibility of the parameters in the two $|t|$ regions can be verified by evaluating
\begin{equation}
\label{eq:chi sq param}
\chi_p^2 = \Delta_p^\T \mat V_p^{-1} \Delta_p\ ,\quad \Delta_p =
\begin{pmatrix}
a_1 - a_2\\
b_1 - b_2\\
\end{pmatrix}\ ,
\end{equation}
where $\mat V_p$ is the covariance matrix for the difference vector $\Delta_p$. It yields $\chi_p^2 = 65.2$ which with $2$ degrees of freedom corresponds to a p-value of $7\cdot10^{-15}$ and a significance of $7.8\un{\sigma}$. This, in turn, rules out the hypothesis of a purely exponential behaviour of the data over the entire observed range.

Since parameters estimated with the least squares method are unbiased, the test in Eq.~(\ref{eq:chi sq param}) is asymptotically binning independent. Indeed, applying it to the data in the ``optimised'' binning (prior to the statistical uncertainty rescaling, Section \ref{sec:stat unc adj}) yields $\chi^2_p = 65.9$ which corresponds to a significance of $7.8\un{\sigma}$. After the uncertainty rescaling, the exclusion significance increases to $8.9\un{\sigma}$.

Finally, it should be emphasised that the above exclusion of purely exponential behaviour is entirely robust against certain systematic errors, most notably normalisation and beam momentum offset. The former can only affect the intercept $\d\sigma/\d t|_{t=0}$, the latter can also scale the parameters $b_i$. However, none of them can bring the parameters $b_2$ and $b_3$ to zero or vice versa.


\section{Conclusions and outlook}
\label{sec:conclusions}
Thanks to a very-high statistics data set TOTEM has excluded a purely exponential differential 
cross-section for elastic proton-proton scattering with significance greater than $7\,\sigma$
in the $|t|$ range from 0.027 to 0.2\,GeV$^{2}$ at $\sqrt{s}=8\,$TeV. The data
are described satisfactorily with an exponent quadratic or cubic in $t$.
Using this refined parametrisation for the extrapolation to the optical point,
$t = 0$, yields total cross-section values compatible with the previous measurement, in all cases
neglecting the effects due to the Coulomb interaction.

In an upcoming analysis, this proof of non-exponentiality in a 
$t$-domain strongly dominated by hadronic interactions will be combined with a measurement of 
elastic scattering in the Coulomb-nuclear interference region,
thus allowing to study the role of the Coulomb interaction in the non-exponential behaviour.


\section*{Acknowledgements}


This work was supported by the institutions listed on the front page and also by the 
Magnus Ehrnrooth foundation (Finland), the Waldemar von Frenckell foundation (Finland), 
the Academy of Finland, the Finnish Academy of Science and Letters (the Vilho, Yrj\"o and Kalle V\"ais\"al\"a Fund), 
the OTKA grant NK 101438 (Hungary). Individuals have received support from Nylands nation vid Helsingfors universitet (Finland) 
and from the M\v SMT \v CR (Czech Republic).


\end{document}